\documentclass{revtex4}
\usepackage{ipe,epsf}
\usepackage{amsmath}
\usepackage{times,mathmine}
\def\Fig#1{Figure \ref{#1}}
\def\Figs#1{Figures \ref{#1}}
\def\Eq#1{Eq.~(\ref{#1})}
\def\Eqs#1{Eqs.~(\ref{#1})}
\def\Ep#1{(\ref{#1})}

\begin{document}

\title{Cracks Cleave Crystals}
\author{M. Marder}
\address{Center for Nonlinear Dynamics and Department of Physics\\
The University of Texas at Austin, Austin TX 78712}

\begin{abstract}
  The problem of finding what direction cracks should move is not
  completely solved. A commonly accepted way to predict crack
  directions is by computing the density of elastic potential energy
  stored well away from the crack tip, and finding a direction of
  crack motion to maximize the consumption of this energy. I provide
  here a specific case where this rule fails. The example is
  of a crack in a crystal. It fractures along a crystal plane, rather
  than in the direction normally predicted to release the most energy.
  Thus, a correct equation of motion for brittle cracks must take into
  account both energy flows that are described in conventional
  continuum theories and details of the environment near the tip that
  are not.
\end{abstract}  

\maketitle 

\section{Introduction}
Much of the continuum theory of fracture concerns itself with the initiation of
cracks, and their speed in response to varying
loads\cite{Broberg.99}. Cracks also choose a direction in which to
move. The continuum theory of fracture has no  law
to describe unambiguously which way they choose.
However, there is a rule that is
widely employed in practice to calculate the direction of crack
motion. This rule is the {\it principle of local symmetry}. It says
that cracks advance in the direction such that shear stresses on the
faces of the crack vanish near the tip; the stresses are purely
tensile, and pull the crack faces apart. Equivalently, cracks move in a
direction that maximizes the consumption of energy stored in
linear elastic fields in front of the tip. The rule was first
proposed for slowly moving cracks by Goldstein and
Salganik\cite{Goldstein.Salganik}, 
generalized to rapidly moving cracks by Adda--Bedia, Arias, Ben Amar
and Lund\cite{Adda-Bedia.99}, and has recently been derived carefully
from a variational principle by Oleaga\cite{Oleaga.01}. 

Experimental checks are not numerous, but they confirm the principle
of local symmetry. They have been carried out in amorphous materials such
as glass where alternatives are
difficult to imagine\cite{Yuse.93,Adda-Bedia.95,Ravi-Chandar.97}.
Experiments in crystals, and the art of cutting gems\cite{Field.71},
find a tendency of cracks to travel along special atomic
planes\cite{Sherman.00,Deegan.03}. However, the crystals where these
experiments have been performed are macroscopically anisotropic.  The
preference of cracks for certain directions can be attributed to the
lack of isotropy in the continuum theory. Thus it has been reasonable
to believe that cracks in a macroscopically homogeneous and isotropic
material should always move in accord with the dictates of local
symmetry.

I will provide here a specific system where the principle of
local symmetry is not obeyed, despite the fact that macroscopically the system is
homogeneous and isotropic.  The demonstration comes from combined
analytical and numerical work. The numerical computations involve
small numbers of atoms (40,000).  However, by combining the
computations with scaling theory, one can predict the outcome of
experiments with arbitrarily large numbers of atoms, and over
arbitrarily large time intervals\cite{Fineberg.99}.

\section{Ideal Brittle Crystal}

\begin{figure}
\epsfxsize\textwidth\epsffile{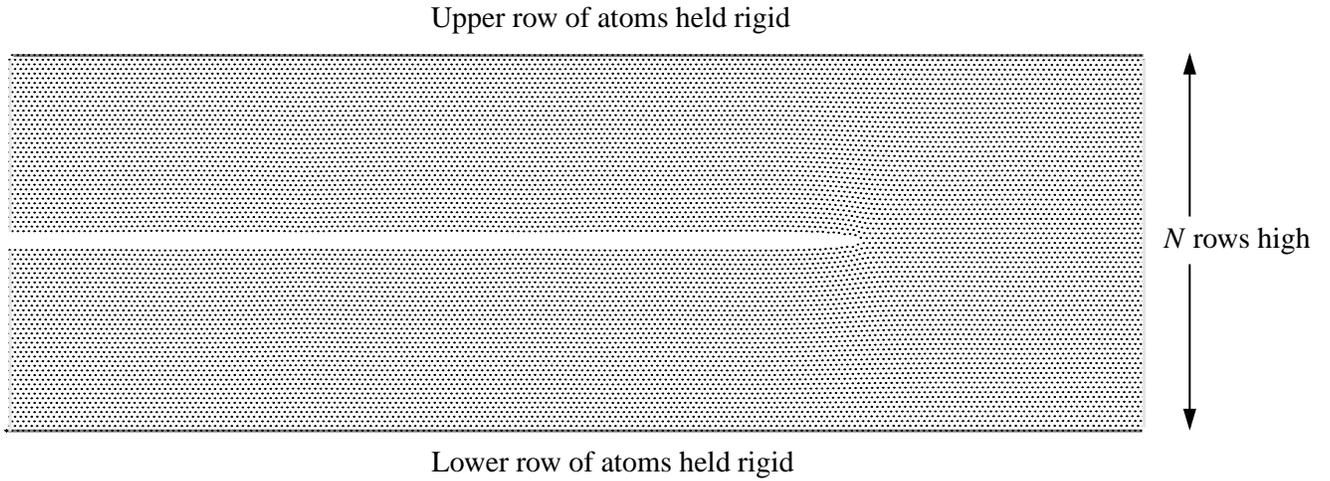}
\caption{Setting for numerical experiment to find steady crack
  states. Atoms are originally arrayed in triangular lattice 80 rows
  high, and three times as long as it is tall. Primitive vectors for
  the equilibrium lattice are $a(1\ 0)$ and $a(1/2\ \sqrt{3}/2)$. The
  crack tip is defined as the location 
  of the rightmost atom whose nearest vertical neighbor is at distance
  greater than $2.5a$. When the crack tip approaches within $60a$ of
  the right boundary, 10 columns of new crystal are attached to the
  right boundary, and the same amount discarded from the left hand
  side. In the discussion leading to   \Eq{eq:Delta}, top
  and bottom rows of atoms are held rigid and stretched vertically apart by a
  distance $\delta_y$. To produce Figure \ref{fig:stt},the top
  boundary is also slid horizontally relative to the bottom by  an
  amount $\delta_x$.}
\label{fig:crack_configuration}
\end{figure}

The material in which I will investigate crack motion is an ideal
brittle triangular crystal with equilibrium lattice spacing $a$ in
which atoms obey the equation of motion 
\begin{equation}
m\ddot {\vec u}_i=\sum_j\left [ \vec f(\vec u_{ji})+\vec g(\dot{\vec
  u}_{ji},{\vec u}_{ji})\right ],
\label{eq:IBC0}
\end{equation}
with
$\vec u_{ji}\equiv\vec u_j-\vec u_i$. 
The functions $\vec f$ and $\vec g$  have the specific forms
\begin{equation}
\vec f(\vec r)=\kappa {\hat r}(r-a)\theta(r_c-r);\ \ 
\vec g(\dot{\vec r},\vec r)= \beta  \dot {\vec r}\,\theta(r_c-r).
\label{eq:IBC1}
\end{equation}
Atoms interact with a central force $f$ that varies linearly around the
equilibrium spacing of length $a$, and whose scale is set by
$\kappa$. If the distance between atoms 
increases to more than $r_c$, the force drops abruptly to zero. In
addition, atoms experience Kelvin dissipation $g$; its scale is set by
$\beta$, is proportional to
the relative velocities of neighbors, and also drops to zero when the
distance between neighbors exceeds $r_c$. The macroscopic elastic
theory of this crystal is homogeneous and isotropic, with Young's
modulus $Y=(5\sqrt{3}/ 4)(\kappa/ a)$ and Poisson ratio $\nu={1/4}$.

I will consider systems that are an even number $N$ atomic planes
high. For cracks to run through such crystals, they must be under
tension. The theory of fracture\cite{Fineberg.99} says that the
tension must be great enough ahead of the tip so that the energy
stored in a vertical 
slice one unit cell wide is enough to snap a pair of bonds at the
crack tip (the Griffith criterion). Bonds snap when they are stretched
beyond their original length by an amount $r_c-a$, so the 
elastic energy stored in a vertical strip of material of horizontal
length $a$ needs to be at least
\begin{equation}
2 \mbox{(two bonds per node)} \times {1\over 2}\times \kappa
(r_c-a)^2=\kappa (r_c-a)^2.
\end{equation}

There are $N-1$ rows of slanted bonds. When they are stretched
vertically  by a small distance $\epsilon_y$, the length of each bond
increases to linear order by an amount 
$\epsilon_r=(\sqrt{3}/ 2)\epsilon_y.$
Therefore, at the Griffith energy, the total energy stored per unit length is
\begin{eqnarray}
&&2 \mbox{(two bonds per site)} \times (N-1)\times {1\over 2}\times
  \kappa (\epsilon_r)^2=\kappa (r_c-a)^2. \\[-1.4cm]
&&\Rightarrow \epsilon_r=(r_c-a)/\sqrt{N-1}
\Rightarrow \epsilon_y= {2\over\sqrt{3}}(r_c-a)
  /\sqrt{N-1}\quad\quad\mbox{\epsfxsize=1.75cm
  \epsffile{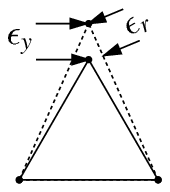}} 
\end{eqnarray}

Then for fracture first to be energetically possible, one rigidly
raises the top of the crystal above its equilibrium position by a
vertical distance $y_c$
\begin{equation}
y_c= {2\over\sqrt{3}}(r_c-a) \sqrt{N-1}.
\label{eq:yc}
\end{equation}

\begin{figure}[!tp]
  \begin{center}
    \leavevmode
    \epsfxsize=3.5in\epsffile{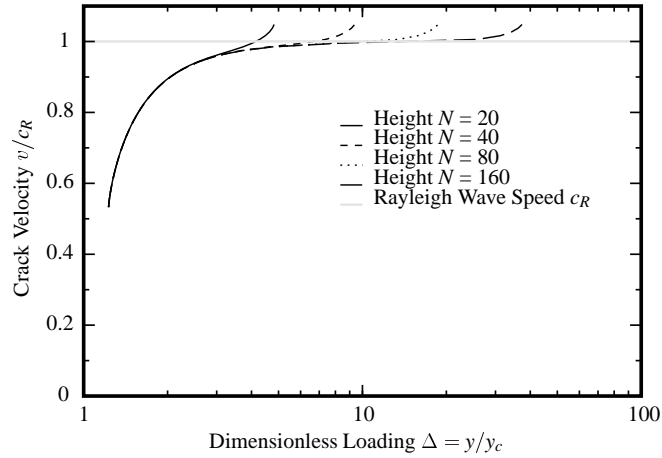}
    \caption{Relationship between velocity $v$ and dimensionless
      loading $\Delta$ for lattice strips of varying height $N$ in
      pure tension. The
      calculations are performed in the limit $r_c\rightarrow a$ with
      Kelvin dissipation $\beta=0.01$ with the Wiener--Hopf technique\cite{Slepyan.81,Slepyan.02,Marder.95.jmps}. The left--hand portions of the
      curve are almost completely independent of system height. The
      cracks are presumed to travel along a weak interface that
      precludes transverse instabilities, and therefore the curves
      continue up and through the Rayleigh wave speed $c_R$. Were the curves
      to be terminated at the points where cracks become unstable in
      homogeneous crystals, they would be nearly indistinguishable. }
    \label{fig:kelvin_scaling}
  \end{center}
\end{figure}
In the limit $r_c\rightarrow a$ steady state cracks in this crystal
are described by exact analytical
solutions\cite{Slepyan.81,Slepyan.02,Kessler.99,Marder.95.jmps}. The most
important observation to extract from these solutions is that the
natural dimensionless measure of how much one has loaded the crystal
is obtained by rigidly displacing its upper surface a distance $\delta_y$ and
then forming the ratio
\begin{equation}
\Delta\equiv{\delta_y/ y_c}.
\label{eq:Delta}
\end{equation}
That is, $\Delta$ is a variable proportional to the strain applied far
ahead of the crack. It equals 1 when the crystal has been loaded
precisely to the Griffith point where fracture first becomes
possible. The analytical solutions demonstrate that if one measures
crack speed $v$ and plots it as a function of loading $\Delta$, the
results become independent of system height $N$ to better than 1\% for
surprisingly small values of $N$, on the order of $N=50$, as shown in
Figure \ref{fig:kelvin_scaling}. This statement is true so long
as the crack speed is not too large. Continuum theory predicts that
cracks in tension cannot exceed the Rayleigh wave speed $c_R$, which for
this model equals $.563 \sqrt{\kappa a^2/ m}$. Figure
\ref{fig:kelvin_scaling} shows that crack speed is
practically independent of system height for $N>50$ and $v<.9c_R$.

The model Eq.~(\ref{eq:IBC0}) is more realistic when
$r_c= 1.2 a$ than in the limit $r_c\rightarrow a$, since
atomic bonds in real brittle materials acutally give way when extended by
about 20\%. For $r_c=1.2$, analytical techniques are no longer
available to provide exact solutions. However, the scaling properties
provided by the analytical solutions continue to hold. The relationship between
crack speed $v$ and loading $\Delta$ is practically independent of system
height $N$ once $N$ reaches a value of around 50. As a result one
can accurately predict the relationship between crack velocity $v$ and
loading $\Delta$ up to the macroscopic limit by performing
computations in systems of microscopic dimensions.

Having established that microscopic computations have a legitimate
macroscopic interpretation, I will now set out to show that the
principle of local symmetry does not always correctly predict crack
paths. Instead there is an interplay between the direction preferred
by far--field stresses, and the direction preferred by microstructure.
This conclusion comes from  seeding cracks on the
centerline of strips as in Figure \ref{fig:crack_configuration}, but then
loading them with a mixture of tension and shear. The top and bottom
boundaries are displaced vertically by distance $\delta_y$ and
horizontally by distance $\delta_x$. 

\begin{figure}[tbp]
  \begin{center}
    \leavevmode
    (A)\epsfxsize=2.5in\epsffile{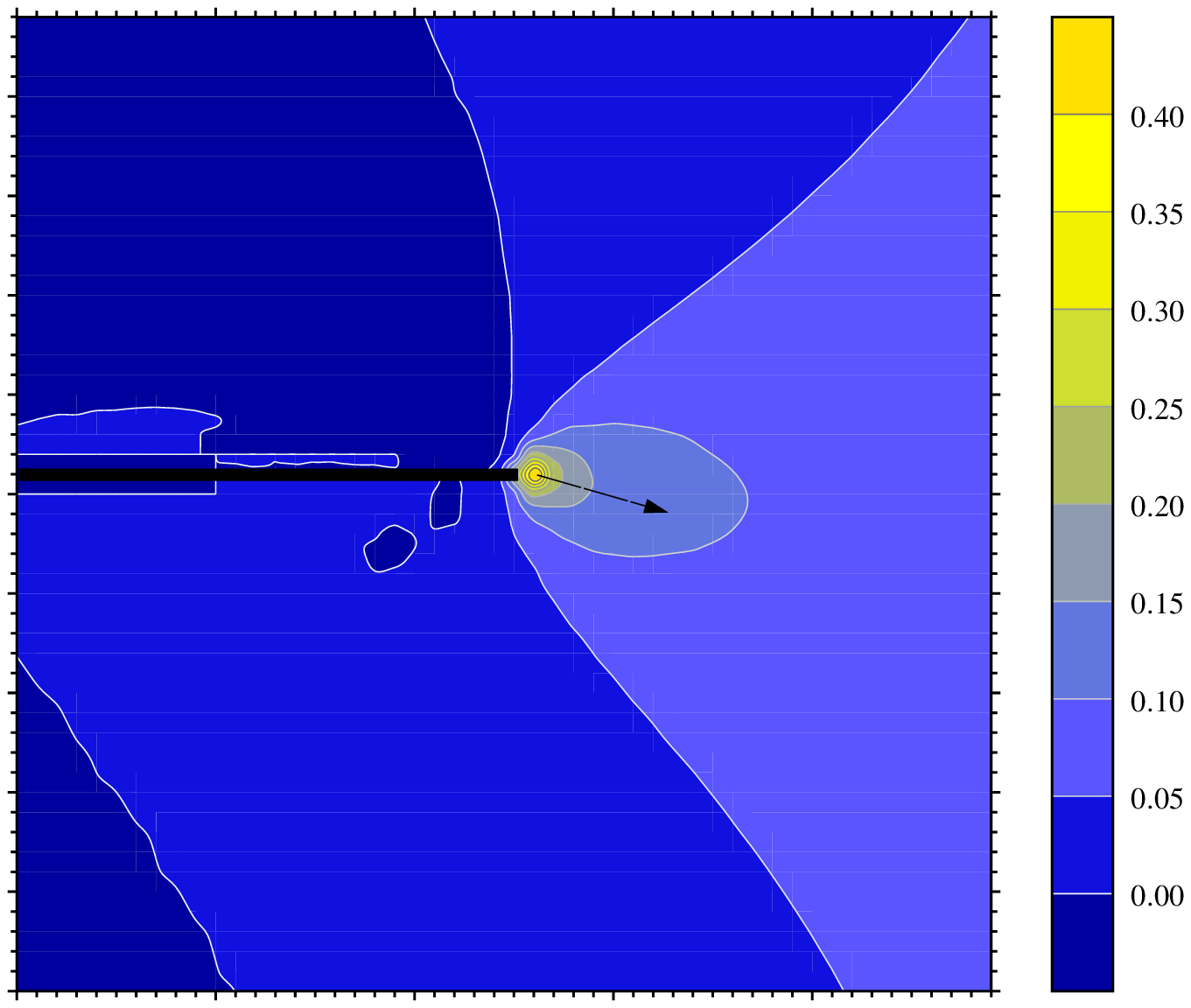}
    (B)\epsfxsize=2.65in\epsffile{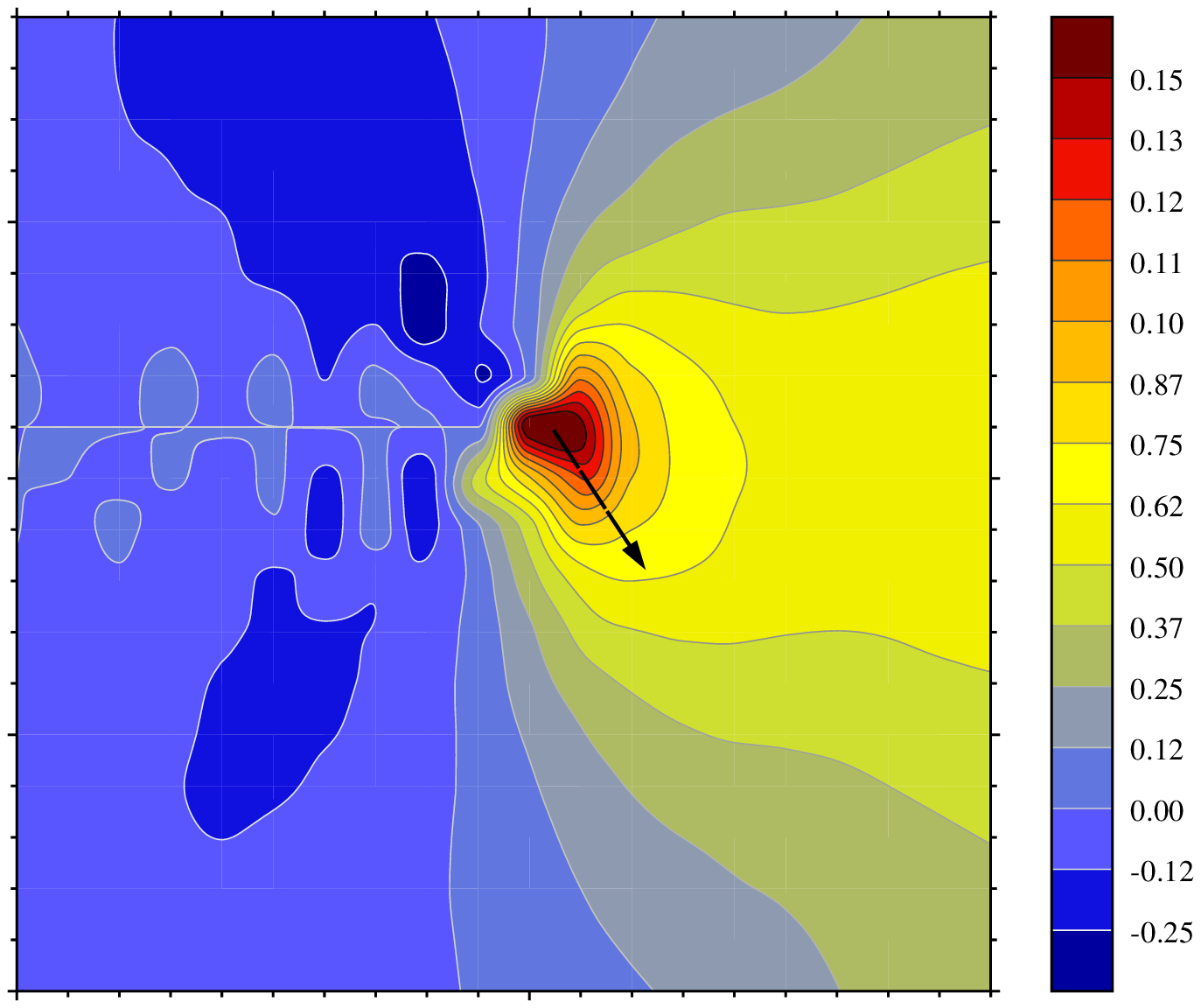}
    \caption{Color contour plots of tensile stress field
    $\sigma_{\theta\theta}$ surrounding 
      crack tips in strips with rigid vertical tensile ($\delta_y$) and
      horizontal shear ($\delta_x$) displacements of upper and lower
      boundaries, computed from solutions of Eq. \ref{eq:IBC0}. Arrows 
      show directions the cracks should turn
      according to local symmetry. Instead, they travel stably
      forever along the horizontal axis. The circular stress islands
      to the left of the crack result from averaging over
      high--frequency waves emitted by the crack and traveling left to
      right. (A) System  $N=150$ rows
      high, $\delta_y/y_c=1.29$, $\delta_x/\delta_y=.23$,  $r_c=1.2$,
      force constant $\kappa=1$, and Kelvin dissipation        $\beta=2$, resulting in a crack
      velocity $v/c_R=.01$. (B) As in (A), but $N=200$ rows high and Kelvin
      dissipation $\beta=.02$, resulting in a crack velocity
      $v/c_R=.83$. The smaller value of $\beta$ is completely
      responsible for the larger crack speed. The larger system is
      chosen because details of the fast--moving crack are more
      difficult to resolve.
 }
    \label{fig:stt}
  \end{center}
\end{figure}

The main result is represented in Figure \ref{fig:stt}. This figure
shows contours of the tensile opening stress surrounding crack tips
moving at two different speeds in a crystal. The contours are tilted
away from the horizontal axis. The principle of local symmetry
predicts that a crack tip surrounded by such stress fields should
rapidly turn and move toward the tip of the largest lobe. However, the
cracks move steadily and stably along the horizontal axis forever.
These cracks follow crystal planes, not external stress fields. It
should be emphasized that the macroscopic elastic properties of a
triangular crystal are completely isotropic. Only the presence of
atomic--scale planes can explain the failure of the cracks to follow
the directions predicted by local symmetry.

\begin{figure}[!tbp]
\begin{center}
\leavevmode
\epsfxsize=2.6in(A) \epsffile{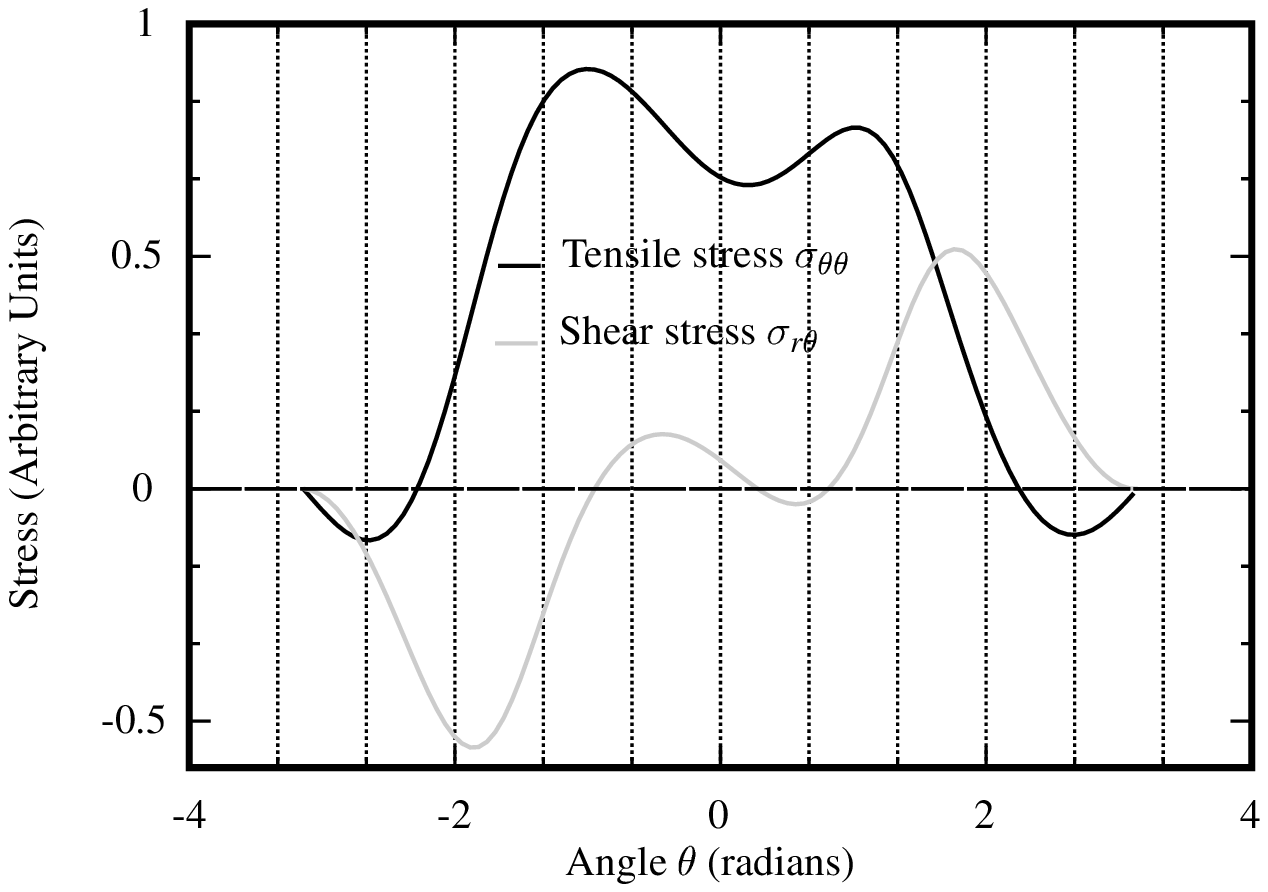}
(B)\epsfxsize=2.6in\epsffile{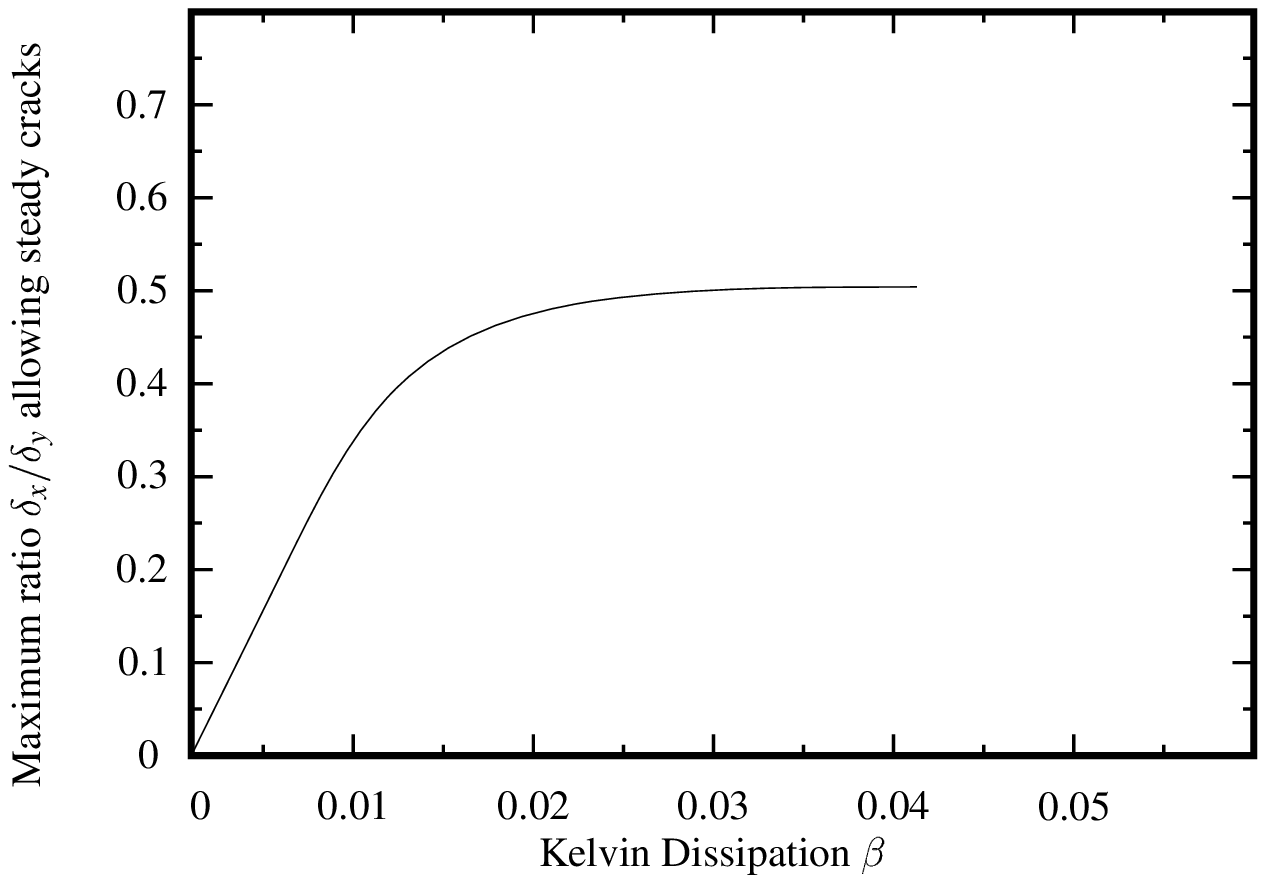} 
\caption{(A) Theoretical asymptotic tensile stress
  $\sigma_{\theta\theta}(\theta)$ and shear stress $\sigma_{r\theta}(\theta)$
  around crack tip for crack traveling at speed $v/c_R=.83$ under 
  the mixed mode loading applied to the crack in
  \Fig{fig:stt}, computed with continuum techniques of
  \cite{Ravi.03}. (B). \label{fig:stt_analytic} (B) Diagram
  showing  combinations of Kelvin dissipation 
      and mixed-mode loading that permit steady motion down center of
      strip from solutions of Eq. \ref{eq:IBC0} . The simulations
  begin with purely tensile loading at $\Delta=1.29$, ($\delta_x=0$)
  and then increase the horizontal displacement of the upper boundary
  $\delta_x$ while keeping the vertical displacement $\delta_y$
  fixed until steady motion becomes unstable.}  
\label{fig:analytics}
\end{center}
\end{figure}

To check the principle of local symmetry, it is necessary to compute
the continuum elastic fields surrounding these cracks.  This task has been
performed in two ways, which agree. First, the elastic stress fields were computed
directly from the positions of atoms in the simulation by taking
binned spatial averages in volumes $V_0$ of
$\sigma_{\alpha\beta}=(1/ 4V_0)\sum_j [ r_{ij}^\alpha
f_{ij}^\beta+ r_{ij}^\beta f_{ij}^\alpha],$
where $\vec f_{ij}$ is the force between atoms $i$ and
$j$\cite{Lutsko.88}. 
Second, the system was viewed as a fracture in a continuous
elastic strip, and the stress fields around the crack tip were 
computed exactly with techniques of fracture mechanics\cite{Ravi.03}
(Figure \ref{fig:stt_analytic}(A)). 

The cracks in \Fig{fig:stt} travel in lattices where the upper
boundary is rigidly displaced by amounts $(\delta_x,\delta_y),$ where
$\delta_x/\delta_y=\tan(.073\pi)$. Fracture mechanics
calculation predicts that tensile stresses are maximal at
angles of $\theta=-16^\circ$ (case (A)) and $\theta=-57^\circ$
(case (B)) to
the $x$ axis, as shown in \Fig{fig:stt_analytic}. 
Therefore, the principle of local symmetry predicts that the crack in
(A) should quickly turn and travel at an angle of $-16^\circ$, and the
crack in (B) should quickly turn along an angle of $-57^\circ$. The
arrows in \Fig{fig:stt} show these directions of crack motion as predicted
by the principle of local symmetry. Stress contours, 
computed directly from spatial averages over interatomic forces,
indeed have lobes in the directions continuum theory predicts. But
the cracks do not move in these directions. Instead, they
travel endlessly along the crystal planes defined by the $x$ axis.

How much shear loading can be placed upon a strip before horizontally
traveling cracks become unstable? An answer to this question is
contained in \Fig{fig:analytics} (B) The results depend upon the
strength of Kelvin dissipation ($\beta$ in \Eq{eq:IBC1}). Without any Kelvin dissipation, cracks
even under purely tensile loading are unstable. As $\beta$ increases,
the range of shear loading cracks can 
withstand while running along a crystal plane also
increases. Physically, Kelvin dissipation has this effect because it
damps the motion of atoms in the vicinity of the tip. A crack can only
begin to depart from the $x$ axis by breaking a first horizontal bond above
the main crack line. Breaking such a bond is easiest when atoms in the
vicinity of the tip oscillate with large amplitude as the tip
passes. Kelvin dissipation damps motion of atoms near the tip, and
makes it more stable.

\section{Continuum Theory Revisited} 

The principle of local symmetry does not explain the results depicted
in \Fig{fig:stt}. However, perhaps a simple modification of the usual rule
could do so.  In a crystal, the energy per area $\Gamma(\theta_c)$ a crack
needs to create new surfaces depends upon the angle $\theta_c$ between the
crack and crystalline planes. In an amorphous material, there should be no
such angular dependence of surface energy. This observation suggests
modifying the principle of local symmetry to choose a crack direction
that maximizes the difference between energy flowing to the crack
tip along $\theta_c$  and energy required along $\theta_c$ to create new
surfaces. 

By returning to the derivations of \cite{Adda-Bedia.99} and
\cite{Oleaga.01} I have found a formal procedure to generalize the
principle of local symmetry in this fashion. This generalization does
capture the tendency of cracks in crystals to move along crystal
planes. However, it does not explain quantitatively the results in
\Figs{fig:stt} and \ref{fig:analytics}. For example, I could not find a way
within continuum theory to include the effects of Kelvin dissipation
$\beta$, since Kelvin dissipation is forbidden in continuum fracture
mechanics\cite{Rice.68}. 

Ordinarily in fracture mechanics, crack motion is determined
by a scalar energy flux $G$, which describes the energy per length a
crack can extract from elastic fields by moving forward unit distance.
Adda-Bedia {\it et al.}\cite{Adda-Bedia.99} and Oleaga\cite{Oleaga.01}
define an energy flux vector $\vec G$. The component of this 
vector parallel to the crack tip is the energy flux $G_\|=G$. The
perpendicular component $G_\bot$ gives the energy that would come to
the crack tip if it could somehow be slid upwards normal to its
current direction.
According to Oleaga\cite{Oleaga.01}, crack motion obeys two rules. Let
a low--energy plane 
lie 
along the $\hat x$ axis at $\theta=0$, and let the current angle of
the crack tip relative to this axis be $\theta_c$. First, the crack
chooses direction and speed so that the energy per area $\Gamma$
needed to create new 
surface equals the energy per area brought to the tip:
\begin{subequations}
\begin{equation}
\Gamma(\theta_c)= G = \vec G\cdot [\cos\theta_c\ \hat x+\sin\theta_c\ \hat y].
\label{eq:OL1}
\end{equation}
Second, the crack direction $\theta_c$ is chosen by the condition that
no other angle $\theta$ provides enough energy for the crack to move:
\begin{equation}
\Gamma(\theta)\geq\vec G\cdot [\cos\theta\ \hat x+\sin\theta\ \hat y].
\label{eq:OL2}
\end{equation}
\label{eq:OL}
\end{subequations}
The energy flux vector $\vec G$
depends upon the crack speed, the current direction of the crack tip
$\theta_c$, and the 
asymptotic stress field approaching the tip, but not upon angle
$\theta$; detailed expressions for each component are
given by Eqs. (4.5) and (4.6) in Ref.~\cite{Oleaga.01}.

\begin{figure}[!tbp]
  \begin{center}
    \leavevmode
\epsfxsize=4.0in \epsffile{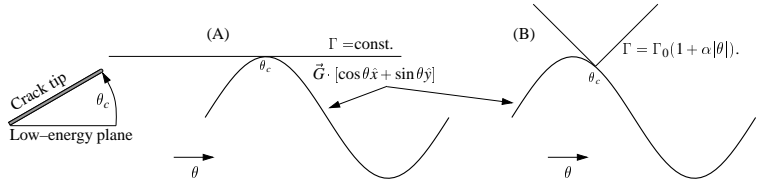}
    \caption{Graphical representations of \Eqs{eq:OL}. Crack directions
      are given by moving the upper curve 
      (representing fracture energy versus angle) down until it first
      touches the lower curve (representing energy flux to crack). The
      contact point of the two curves gives the direction of crack motion
      $\theta_c$. (A) When fracture energy is
      independent of direction, contact can only occur at the maximum of
      the lower curve. This means that $\vec
       G$ is parallel to the direction of crack motion $\theta_c$. (B)
      If fracture energy has a cusp--like minimum, there are many
      ways to satisfy \Eqs{eq:OL}. For some, the upper tip just
      touches the lower curve; for these, the crack  cleaves a plane
      at $\theta_c=0$. For others, $\theta_c$ varies continuously away
      from $0$.  A
      similar graphical construction, suggested by experimental data,
      is found in \cite{Deegan.03}.}
    \label{fig:crack_direction}
  \end{center}
\end{figure}

When the fracture energy $\Gamma$ is independent of direction, these
rules imply the familiar principle of local symmetry. To see why, note that
if $\vec G$ does not point along $\theta_c$
there must be some value of $\theta$ that
will violate the inequality in \Ep{eq:OL2}, as shown in
\Fig{fig:crack_direction} (A). However, if 
$\Gamma(\theta)$ is not constant, matters are not so simple.
In particular, suppose that
$\Gamma$ has the dependence expected for small angles $\theta$ in the
presence of crystal planes 
$\Gamma(\theta)=\Gamma_0(1+\alpha|\theta|),$
where $\alpha>0$ is a constant of order unity and $\Gamma_0$ is the
fracture energy along the plane. 
\Eqs{eq:OL} now have the graphical interpretation shown in
\Fig{fig:crack_direction} (B). 
The crack direction is determined by sliding a sharp tip
over a sine curve. The tip must just touch the sine curve at one
point, and can never dip below it. There
are two sorts of solutions. In the first, contact occurs right at the
tip where $\theta_c=0$, and the crack cleaves the low--energy
plane. Such solutions can occur even if $\vec G$ does 
not point along $\theta=0$, thus violating the customary principle of
local symmetry. A second sort of solution is possible if the slope of
the sine curve somewhere becomes larger than $\alpha$; for these, the
crack can travel at an angle different from $\theta=0$.

How well does this generalized principle of local symmetry work?
\emph{Qualitatively}, it has many sensible features. It correctly predicts
that cracks can be trapped on crystal planes, and run along them for a
range of loading conditions. It predicts that beyond a critical
loading, the crack will stop following the crystal plane.
\emph{Quantitatively}, I have been unable to reproduce the results of
microscopic calculations using \Eqs{eq:OL}. In particular, \Eqs{eq:OL}
predict that \emph{all} the cracks described in \Fig{fig:analytics} (B)
should travel stably along the $x$ axis, and misses the correct result
which is that at a certain value of shear loading these cracks are no
longer stable.

Therefore, I do not
believe there is a substitute for detailed microscopic analysis if one wants to
treat correctly the dynamics of cracks in brittle materials.

\acknowledgements

The National Science Foundation (DMR-9877044 and DMR-0101030)
supported this work. Matt Lane provided many comments to help me
improve this manuscript.

\bibliographystyle{ieeetr}
\bibliography{/home/marder/crack/tex/fracture.bib,/home/marder/grants/marder.bib}

\end{document}